\documentclass[conference]{IEEEtran}

\usepackage{amsmath,amssymb,amsfonts}
\usepackage{stfloats}
\usepackage{cite}
\usepackage{graphicx}
\usepackage{psfrag}
\usepackage{subfigure}
\usepackage{amsmath}
\usepackage{array}
\usepackage{algorithmicx}
\usepackage{algpseudocode}
\usepackage{setspace}
\usepackage{blindtext}
\usepackage{url}
\usepackage{epstopdf}
\usepackage{verbatim}
\usepackage{color}
\usepackage{amsmath}
\usepackage{bm}
\usepackage{multirow}
\usepackage{booktabs}
\usepackage{makecell}
\usepackage{ntheorem}
\usepackage{textcomp}
\usepackage{amsmath}
\def\BibTeX{{\rm B\kern-.05em{\sc i\kern-.025em b}\kern-.08em T\kern-.1667em\lower.7ex\hbox{E}\kern-.125emX}}

\newtheorem{Prob}{Problem}

\title{Joint Visual and Wireless Signal Feature based Approach for High-Precision Indoor Localization}
\author{Yu Wang$^{\dagger}$, Guangbing Zhou$^{\dagger}$ $^{\ddag}$, Chenlu Xiang$^{\dagger}$, Shunqing Zhang$^{\dagger}$, and Shugong Xu$^{\dagger}$ \\
$^{\dagger}$ Shanghai Institute for Advanced Communication and Data Science, \\
Key laboratory of Specialty Fiber Optics and Optical Access Networks, \\
Joint International Research Laboratory of Specialty Fiber Optics and Advanced Communication, \\
Shanghai University, Shanghai, 200444, China\\
$^{\ddag}$ South China Robotics Innovation Research Institute, Guangdong, 528315, China\\
Email: \{wangyu0423, zhou020, xcl, shunqing, shugong\}@shu.edu.cn}

\begin{document}
\maketitle

\begin{abstract}
The existing localization systems for indoor applications basically rely on wireless signal. With the massive deployment of low-cost cameras, the visual image based localization become attractive as well. However, in the existing literature, the hybrid visual and wireless approaches simply combine the above schemes in a straight forward manner, and fail to explore the interactions between them. In this paper, we propose a joint visual and wireless signal feature based approach for high-precision indoor localization system. In this joint scheme, WiFi signals are utilized to compute the coarse area with likelihood probability and visual images are used to fine-tune the localization result. Based on the numerical results, we show that the proposed scheme can achieve 0.62m localization accuracy with near real-time running time.
\end{abstract}


\section{Introduction} \label{sect:intro}
The off-the-shelf localization systems installed in modern smartphones or Internet-of-Things (IoT) devices, mainly rely on analyzing real-time wireless signal features \cite{hofmann2007gnss,faragher2015location,zhang2019fingerprint,xiang2019robust}, where \emph{global navigation satellite system} (GNSS) signals \cite{hofmann2007gnss} are often utilized for outdoor scenarios and short range wireless signals are selected for indoor cases. Together with \emph{inertial navigation system} (INS) \cite{groves2015principles}, the GNSS based outdoor solution is able to reach meter level seamless localization accuracy, while it is still quite challenging to achieve the same level in the indoor environment using either Bluetooth low energy (BLE) \cite{faragher2015location} or widely deployed 3GPP LTE/5G and WiFi signals \cite{zhang2019fingerprint,xiang2019robust}.

With the massive deployment of low-cost digital cameras in the smart entities, a potential approach to further improve the localization accuracy is to incorporate the visual information, which is often referred to as visual based localization (VBL) \cite{piasco2018survey}. Although the conventional VBL design is referring to a classical camera pose estimation task containing both position and orientation detection, the corresponding applications are still limited to computer vision areas, such as simultaneous localization and mapping (SLAM) \cite{durrant2006simultaneous} or structure from motion (SfM) \cite{schonberger2016structure}, and the current localization accuracy in the indoor environment is around one-half meter for $2.5m \times 2m \times 1.5m$ office areas as reported in \cite{kendall2015posenet}.

Since the signal feature based localization (SFBL) and the VBL schemes share the similar design philosophy to establish databases offline and perform pattern matching online, a natural extension is to jointly minimize the localization errors via combining databases and matching algorithms in a brute force manner and the localization performance can be improved to meter level \cite{jiao2018smart}, or sub-meter level with the help of high-cost lidar. A smarter approach is to decouple SFBL and VBL processing in a hierarchical way \cite{dong2015imoon}, e.g., to perform coarse-grained localization using SFBL and fine-tune the intermediate positions using VBL, and the resultant errors can be reduced to less than 2 m. This approach requires significant manpower for image database generation and online processing complexity. To make it more practical, \cite{xu2016indoor} proposes to project the query images into two dimensional floor plan for pattern matching and \cite{hu2017wi} utilizes the special `EXIT' signs to reduce the processing complexities.

The above hybrid SFBL and VBL approaches provide a promising direction for high-precision indoor localization by utilizing the corresponding advantages in a separated manner. However, it fails to explore the interactions between two schemes, and the coarse localization results from SFBL, as shown later, can be incorporated in the later VBL procedures to reduce the localization errors as well as the processing complexities. In this paper, we propose a joint visual and wireless signal feature based solution for high-precision indoor localization, where the main contributions are listed below.
\begin{itemize}
\item{\bf Signal Feature Assisted VBL.} The conventional hybrid SFBL and VBL scheme simply select some candidate regions using SFBL to restrict the processing complexities in the VBL stage. If we regard the region index identification as `hard decision', a more reasonable scheme is to consider `soft decision' instead. Hence, in this paper, we propose a joint visual and wireless signal feature based localization (JVWL) by considering the likelihood distribution of potential positions, which eventually help to improve the localization accuracy.
\item{\bf Fast Sign-free Detection.} With the obtained likelihood distribution of potential positions, our proposed JVWL framework, including a CNN based pre-trained feature extraction block and a low order recurrent architecture for location prediction, is able to understand the inner relationships between the stochastic wireless signal knowledge and the deterministic visual information. By combining two types of features together, our proposed scheme is able to figure out in-depth characteristics, which are previously offered by special signs.
\item{\bf Low Complexity Database Construction.} In addition, we propose a low complexity database construction mechanism on top of the JVWL framework, which uses the visual images of near-by indoor environment to infer the remaining parts alongside corridors.
\end{itemize}

The rest of paper is organized as follows. The entire localization system is described in Section~\ref{sect:sys} and the proposed joint localization scheme is discussed in Section~\ref{sect:sol}. In Section~\ref{sect:exper}, we present our experimental results and the concluding remarks are provided in Section~\ref{sect:conc}.

\section{System Model}\label{sect:sys}
In this section, we introduce the overall procedures of the proposed JVWL scheme and discuss the construction of databases in what follows.

\begin{figure}
\centering
\includegraphics[width = 3 in]{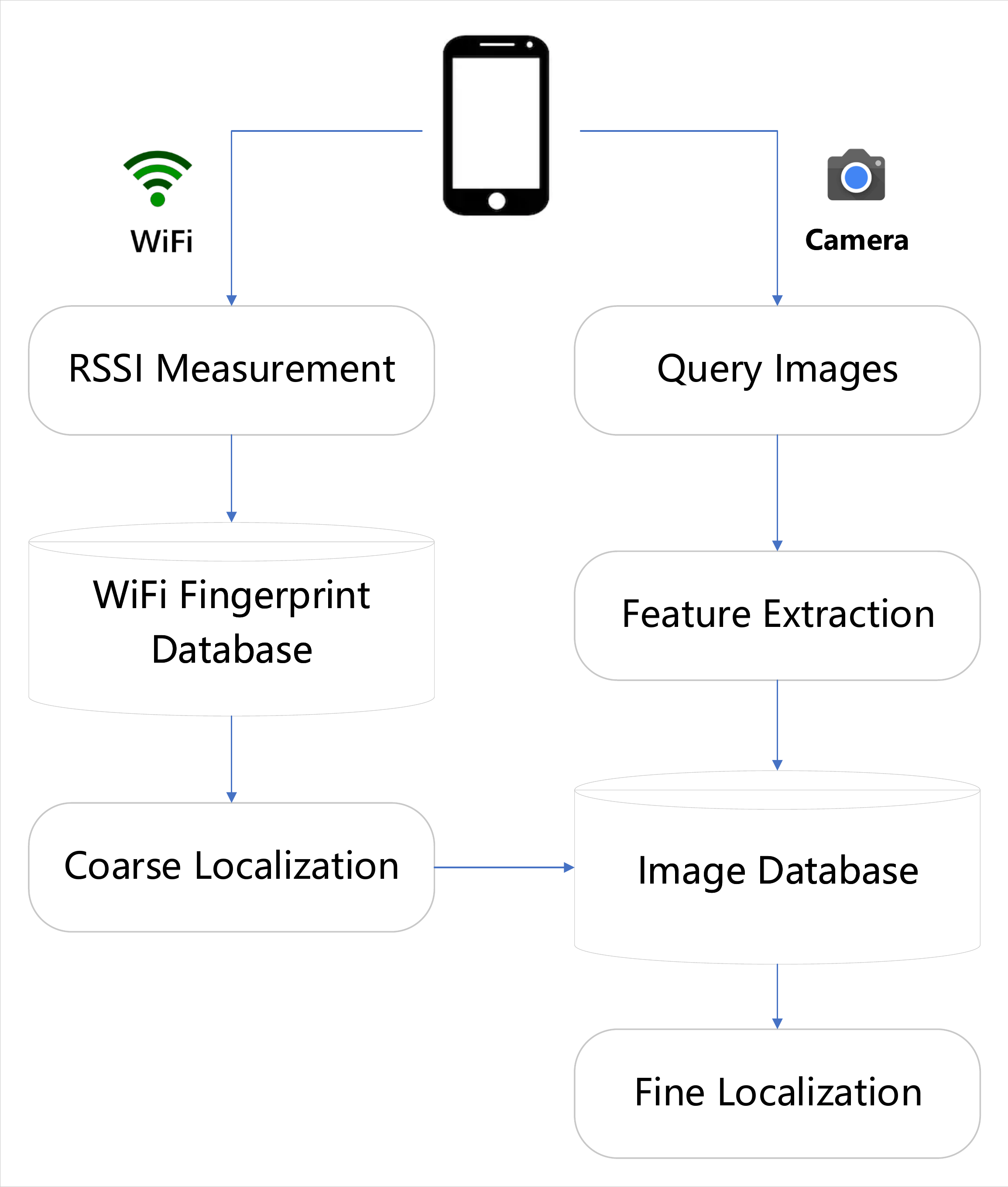}
\caption{The overall procedures of the proposed JVWL scheme. It contains two stages, including WiFi RSSI based coarse localization and visual image based fine localization.}
\label{fig:system}
\end{figure}

\subsection{Overall Description}
As shown in Fig.~\ref{fig:system}, the proposed JVWL scheme first collects $N_s$ sample received signal strength indications (RSSIs) from $N_{AP}$ WiFi access points, and $N_p$ query images with $N_{w} \times N_{l}$ pixels and $N_{RGB}$ color channels from on-device cameras, where the corresponding observations are denoted as $\mathbf{R}(\mathcal{L}) \in \mathbb{R}^{N_{s} \times N_{\textrm{AP}}}$ and $\mathbf{I}(\mathcal{L}) \in \mathbb{Z}^{N_{p} \times N_{w} \times N_{l} \times N_{RGB}}$ for a given location $\mathcal{L}$, respectively. RSSIs of $N_{RP}$ reference points (RPs) are offline collected to construct WiFi fingerprint database, $\mathcal{DB}_{W}$, which consists of $N_{RP}$ RP locations, $\{\mathcal{L}^{i}_{RP}\}$, and the measured RSSIs, $\{\mathbf{R}(\mathcal{L}^{i}_{RP})\}$. The image database, $\mathcal{DB}_{I}$, is constructed in a similar manner, which contains the location $\mathcal{L_I}$, and the associated images, $\{\mathbf{I}(\mathcal{L_I})\}$.

\begin{figure}
\centering
\includegraphics[width = 3 in]{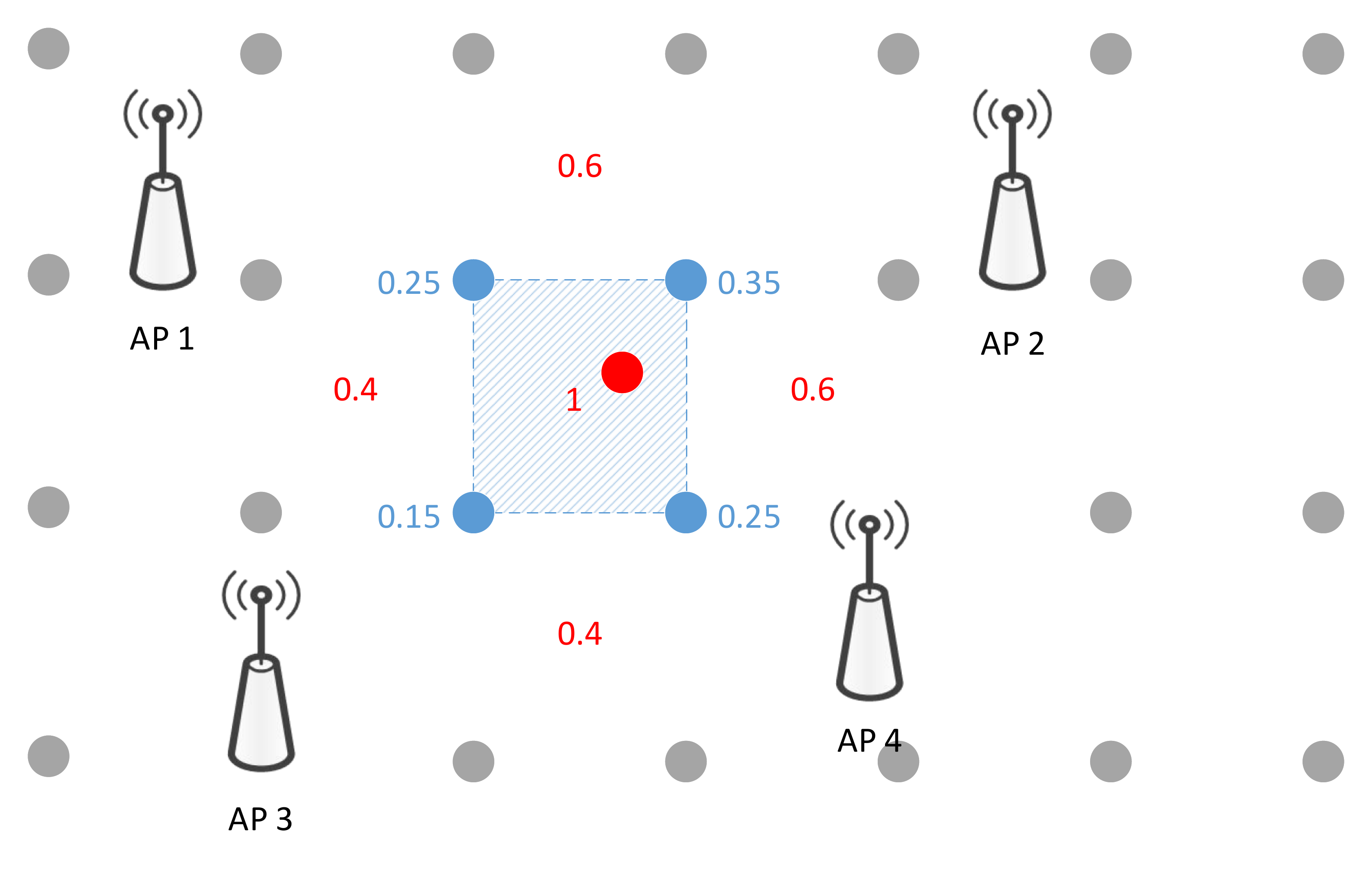}
\caption{The illustration of coarse localization. The nearest $N_{p}$ RPs are selected by KNN algorithm and form a circle area, which is the coarse localization area. The gray dots and red dot are RP positions and test position respectively.}
\label{fig:coarse}
\end{figure}

The entire localization procedures consist of a WiFi RSSI based coarse localization and a visual image based fine localization\footnote{Kindly note that the above RSSI and visual information can be easily obtained via existing smartphone sensors, such as WiFi receivers and cameras.}, as explained below.
\begin{itemize}
    \item {\em Coarse Localization $f(\cdot)$} In the coarse localization, the proposed JVWL scheme first computes the likelihood probability with respect to (w.r.t.) $N_{RP}$ RPs, i.e., $\hat{\mathbf{p}}_{RP}(\mathcal{L}) = [\hat{p}^{1}_{RP}(\mathcal{L}), \ldots, \hat{p}^{i}_{RP}(\mathcal{L}), \ldots, \hat{p}^{N_{RP}}_{RP}(\mathcal{L})]$, by inquiring the WiFi fingerprint database, $\mathcal{DB}_{W}$. By comparing with the observed WiFi RSSIs, the likelihood probability w.r.t. the $i^{th}$ RP can be obtained via a standard support vector machine (SVM) scheme, which is proven effective in fingerprint classification tasks \cite{zhou2017device}, e.g.,
    \begin{eqnarray}
    \hat{p}^{i}_{RP}(\mathcal{L}) = f_1 \left(\mathbf{R}(\mathcal{L}), \mathcal{DB}_{W}\right).\label{eqn:prob_rp}
    \end{eqnarray}
    In order to reduce the searching complexity of the latter fine localization, we partition the target areas into $N_{\mathcal{A}}$ consecutive areas based on $N_{RP}$ RPs, $\left\{\mathcal{A}^{j}\left(\{\mathcal{L}^{i}_{RP}\}\right)\right\}$, as shown in Fig.~\ref{fig:coarse}. The likelihood probabilities of $N_{\mathcal{A}}$ areas, $\hat{\mathbf{p}}_{\mathcal{A}}(\mathcal{L}) = [\hat{p}^{1}_{\mathcal{A}}(\mathcal{L}), \ldots, \hat{p}^{j}_{\mathcal{A}}(\mathcal{L}), \ldots, \hat{p}^{N_{\mathcal{A}}}_{\mathcal{A}}(\mathcal{L})]$, can be calculated by summing over the likelihood probabilities of RPs, where each element $\hat{p}^{j}_{\mathcal{A}}(\mathcal{L})$ is given by,
    \begin{eqnarray}
    \hat{p}^{j}_{\mathcal{A}}(\mathcal{L}) = f_2\left(\hat{\mathbf{p}}_{RP}(\mathcal{L})\right) = \sum_{i \in \mathcal{A}^{j}\left(\{\mathcal{L}^{i}_{RP}\}\right)} \hat{p}^{i}_{RP} (\mathcal{L}). \label{eqn:prob_a}
    \end{eqnarray}
    The coarse localization result is thus given by selecting $J^{\star}$ most possible areas according to the likelihood probabilities, $\hat{\mathbf{p}}_{\mathcal{A}}(\mathcal{L})$. Mathematically, if we denote $\Omega^{\star}_j(\mathcal{L})$ and $\overline{\Omega^{\star}_j(\mathcal{L})}$ to be the index set of selected areas and its complementary set, the candidate localization area $\mathcal{A}^{\star}\left(\mathcal{L}\right)$ and the corresponding likelihood probability $\hat{\mathbf{p}}_{\mathcal{A}^{\star}}(\mathcal{L})$ can be expressed as,
    \begin{eqnarray}
    \mathcal{A}^{\star}\left(\mathcal{L}\right) & = & \bigcup_{j \in \Omega^{\star}_j(\mathcal{L})} \mathcal{A}^{j}\left(\{\mathcal{L}^{i}_{RP}\} \right), \label{eqn:Area} \\
    \hat{\mathbf{p}}_{\mathcal{A}^{\star}}(\mathcal{L}) & = &  \{\hat{p}^{j}_{\mathcal{A}}(\mathcal{L})\}, \ \forall j \in \Omega^{\star}_j, \label{eqn:Prob}
    \end{eqnarray}
    where $\hat{p}^{j}_{\mathcal{A}}(\mathcal{L}) \geq \hat{p}^{j'}_{\mathcal{A}}(\mathcal{L})$ for any $j \in \Omega^{\star}_j(\mathcal{L})$ and $j' \in \overline{\Omega^{\star}_j(\mathcal{L})}$ and the cardinality of $\Omega^{\star}_j(\mathcal{L})$ is $J^{\star}$. By cascading \eqref{eqn:prob_rp}-\eqref{eqn:Prob}, we denote the entire coarse localization process as,
    \begin{eqnarray}
    \left(\mathcal{A}^{\star}\left(\mathcal{L}\right),\hat{\mathbf{p}}_{\mathcal{A}^{\star}}(\mathcal{L})\right)  = f\left(\mathbf{R}(\mathcal{L}), \mathcal{DB}_{W}, \left\{\mathcal{A}^{j}\left(\{\mathcal{L}^{i}_{RP}\}\right)\right\} \right).\nonumber
    \end{eqnarray}
    \item {\em Fine Localization $g(\cdot)$} In the fine localization, the proposed JVWL scheme maps the $N_{p}$ query images, $\mathbf{I}(\mathcal{L})$, to the estimated location, $\hat{\mathcal{L}} \in \mathcal{A}^{\star}\left(\mathcal{L}\right)$, according to the image database $\mathcal{DB}_{I}$. To control the searching complexity, we only use a subset of the entire image database in the practical deployment, e.g., $\mathcal{DB}_{I}\left(\mathcal{A}^{\star}\left(\mathcal{L}\right)\right) \triangleq \left\{\mathbf{I}(\mathcal{L}_I), \forall \mathcal{L}_I \in \mathcal{A}^{\star}\left(\mathcal{L}\right) \right\} \subset \mathcal{DB}_{I}$, and the mathematical expression of the fine localization process is given by,
    \begin{eqnarray}
    \hat{\mathcal{L}} = g\big(\mathbf{I}(\mathcal{L}), \mathcal{DB}_{I}\left(\mathcal{A}^{\star}\left(\mathcal{L}\right)\right) \big).
    \end{eqnarray}
\end{itemize}

\subsection{Database Construction}
In order to construct the databases $\mathcal{DB}_{W}$ and $\mathcal{DB}_{I}$, a site survey of WiFi RSSI fingerprints and camera images is conducted with $N_{RP}$ and $|\mathcal{L}_I|$ positions, respectively. To make a more reliable database, $N_{W}$ rounds of RSSIs collections and $N_{IR}$ rounds of images will be performed to construct $\mathcal{DB}_{W}$ and $\mathcal{DB}_{I}$, e.g., $\mathcal{DB}_{W} = \left\{ \left(\mathbf{R}^{N_W}(\mathcal{L}^{i}_{RP}),\mathcal{L}^{i}_{RP}\right)\right\}$ and $\mathcal{DB}_{I} = \left\{ \left(\mathbf{I}^{N_I}(\mathcal{L}_{I}),\mathcal{L}_{I}\right)\right\}$, respectively. Since $\mathcal{DB}_{I}\left(\mathcal{A}^{\star}\left(\mathcal{L}\right)\right)$ is equal to $\cup_{j \in \Omega^{\star}_j(\mathcal{L})} \mathcal{DB}_{I}\left(\mathcal{A}^{j}\left(\{\mathcal{L}^{i}_{RP}\}\right)\right)$, we can partition the image database $\mathcal{DB}_{I}$ into $N_{\mathcal{A}}$ parts in the offline stage, e.g., $\left\{\mathcal{DB}_{I} \left( \mathcal{A}^{j}\left(\{\mathcal{L}^{i}_{RP}\}\right)\right)\right\}$, and efficiently construct $\mathcal{DB}_{I}\left(\mathcal{A}^{\star}\left(\mathcal{L}\right)\right)$ in the online stage. For the convenience of data collection and future update, a mobile robot equipped with WiFi, camera, lidar and IMU sensors are used to construct $\mathcal{DB}_{W}$ and $\mathcal{DB}_{I}$ with corresponding ground truth positions. More implementation details are presented in Section~\ref{sect:exper}.

\section{Proposed JVWL Scheme} \label{sect:sol}
In this section, we describe the formulation and design of the proposed JVWL scheme. To be more specific, we propose a joint optimization framework for visual and wireless localization, based on which a novel neural network structure and loss function are then presented.

\subsection{Problem Formulation}
In order to obtain a reliable localization error performance, we introduce the subscript $m$ to the ground-true and estimated locations, and formulate the joint localization problem as follows.

\begin{Prob}[Joint Localization]
\label{prob:joint}
\begin{eqnarray}
\underset{g (\cdot), \{\hat{\mathcal{L}}_m\}}{\textrm{minimize}} && \frac{1}{M} \sum_{m=1}^{M} \min_{j \in \Omega_j^{\star}(\hat{\mathcal{L}}_m)} \frac{\|\hat{\mathcal{L}}_m - \mathcal{L}_{m}\|^2_2}{\hat{p}^{j}_{\mathcal{A}}(\hat{\mathcal{L}}_m)} \\
\textrm{subject to}
&& \left(\mathcal{A}^{\star}\left(\hat{\mathcal{L}}_m\right),\hat{\mathbf{p}}_{\mathcal{A}^{\star}}(\hat{\mathcal{L}}_m)\right)  = f\big(\mathbf{R}(\hat{\mathcal{L}}_m), \nonumber \\ && \mathcal{DB}_{W},  \left\{\mathcal{A}^{j}\left(\{\mathcal{L}^{i}_{RP}\}\right)\right\} \big), \\
&& \hat{\mathcal{L}}_m = g \left(\mathbf{I}(\mathcal{L}_m), \mathcal{DB}_{I}\left(\mathcal{A}^{\star}\left(\hat{\mathcal{L}}_m\right)\right) \right), \\
&& \hat{\mathcal{L}}_m \in \mathcal{A}^{\star}\left(\hat{\mathcal{L}}_m\right), \forall m,
\end{eqnarray}
where $M$ is the total number of localization tasks and $\|\cdot\|_2$ represents the vector $l_2$ norm as defined in \cite{boyd2004convex}, i.e., $\forall x = [x_1, x_2, \ldots, x_K ] \in \mathbb{R}^{K}, \|x\|_2=(\sum_{k=1}^{K}|x_k|^2)^{\frac{1}{2}}$.
\end{Prob}

In the joint formulation of Problem~\ref{prob:joint}, the function $g(\cdot)$ is usually modelled as a typical regression problem according to the existing literature. Compared with the conventional visual based localization technologies, the proposed scheme only needs to evaluated over the potential area $\mathcal{A}^{\star}\left(\hat{\mathcal{L}}_m\right)$, which undoubtedly reduce the computational complexity. More importantly, the high similarity of some indoor scenes will make conventional visual localization unable to distinguish the true area, which will cause additional positioning errors. Therefore the partitioned area $\mathcal{A}^{\star}\left(\hat{\mathcal{L}}_m\right)$ can help to improve the positioning accuracy as well.

 \subsection{Neural Network}
\begin{figure*}
\centering
\includegraphics[width = 6 in]{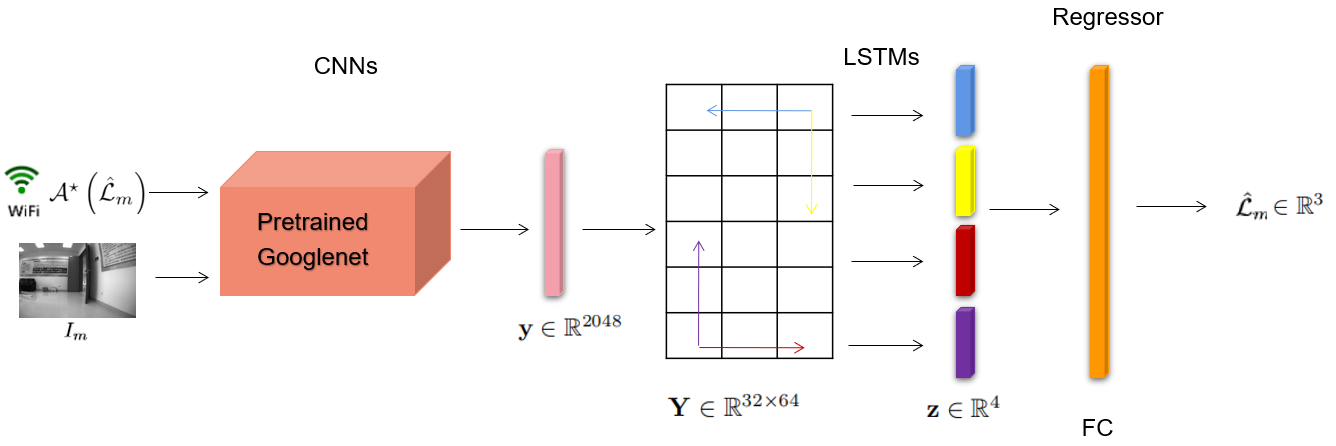}
\caption{The proposed network for fine localization with CNNs, LSTMs and Regressor.}
\label{fig:network}
\end{figure*}

To better describe the function $g(\cdot)$, we design a deep neural network architecture, which is is illustrated in Fig.~\ref{fig:network}.
All the input images are pre-processed into zero mean values and the same pixel size with random central point. The state of the art  GoogLeNet \cite{szegedy2015going} is leveraged as a basis for feature extraction. The pretrained weights from dataset like ImageNet \cite{Deng2009ImageNet} or Places \cite{zhou2014learning}, can improve the representation performance and save the training time. At the end of the CNN layers, a fully connected layer performs average pooling and output a one dimensional feature vector. To further exploit the spatial dependcies of images taken by cameras and improve the localization accuracy, we apply the long short-term memory (LSTM) units in the up, down, left and right directions \cite{inproceedings}, which is capable of memorizing sequential data more quickly. At last, the outputs of LSTM units are then combined and fed to the regression module to predict the target position $\hat{\mathcal{L}}_m$. The detailed configurations and parameters for our neural network are listed in Table~\ref{tab:network}\footnote{Note that each convolutional layer in the CNNs part corresponds 'Conv-ReLU-Max pooling' sequence. Each Inception layer is divided into four branches, which are processed by convolution kernels of different scales.}.
\begin{table}[!ht]
\centering
\caption{An Overview of Network Configuration and Parameters.}\label{tab:network}
\begin{tabular}{ c c c}
\hline
\textbf{Module} & \textbf{Layers} & \textbf{Parameters} \\
\hline
\multirow{2}*{CNNs} & conv$-1$ & $112\times112\times 64$ \\
& conv$-2$ & $56 \times56\times192$  \\
& Inception 1 & $28 \times 28 \times 256$  \\
& Inception 2 & $28 \times 28 \times 480$  \\
& Inception $3-5$ & $14\times14\times512$  \\
& Inception 6 & $14\times14\times528$  \\
& Inception 7 & $14\times14\times832$  \\
& Inception 8 & $7\times7\times832$  \\
& Inception 9 & $7\times7\times1024$  \\
& avg pool  & $1\times1\times1024$  \\
& FC & $1\times1\times2048$  \\
\hline
\multirow{2}*{LSTMs} & reshape & $32\times64$ \\
& lstm & $1\times1\times4$  \\
\hline
Regressor & FC & $3\times1$ \\
\hline
\end{tabular}
\end{table}

For the regression problem, the fully connected (FC) layer is used at the output layer. We design the loss function $\mathbb{L}$ of the neural network according to the previous problem formulation, which is given by,
\begin{eqnarray}
\mathbb{L} =  \min_{j \in \Omega_j^{\star}(\hat{\mathcal{L}}_m)} \frac{\|\hat{\mathcal{L}}_m - \mathcal{L}_{m}\|^2_2}{\hat{p}^{j}_{\mathcal{A}}(\hat{\mathcal{L}}_m)}.
\end{eqnarray}
In addition, the ADAM solver \cite{kingma2014adam} is used for optimization and the learning rate is set as 0.0001.

\section{Experiment Results} \label{sect:exper}
\begin{figure}
\centering
\includegraphics[width = 3.4 in]{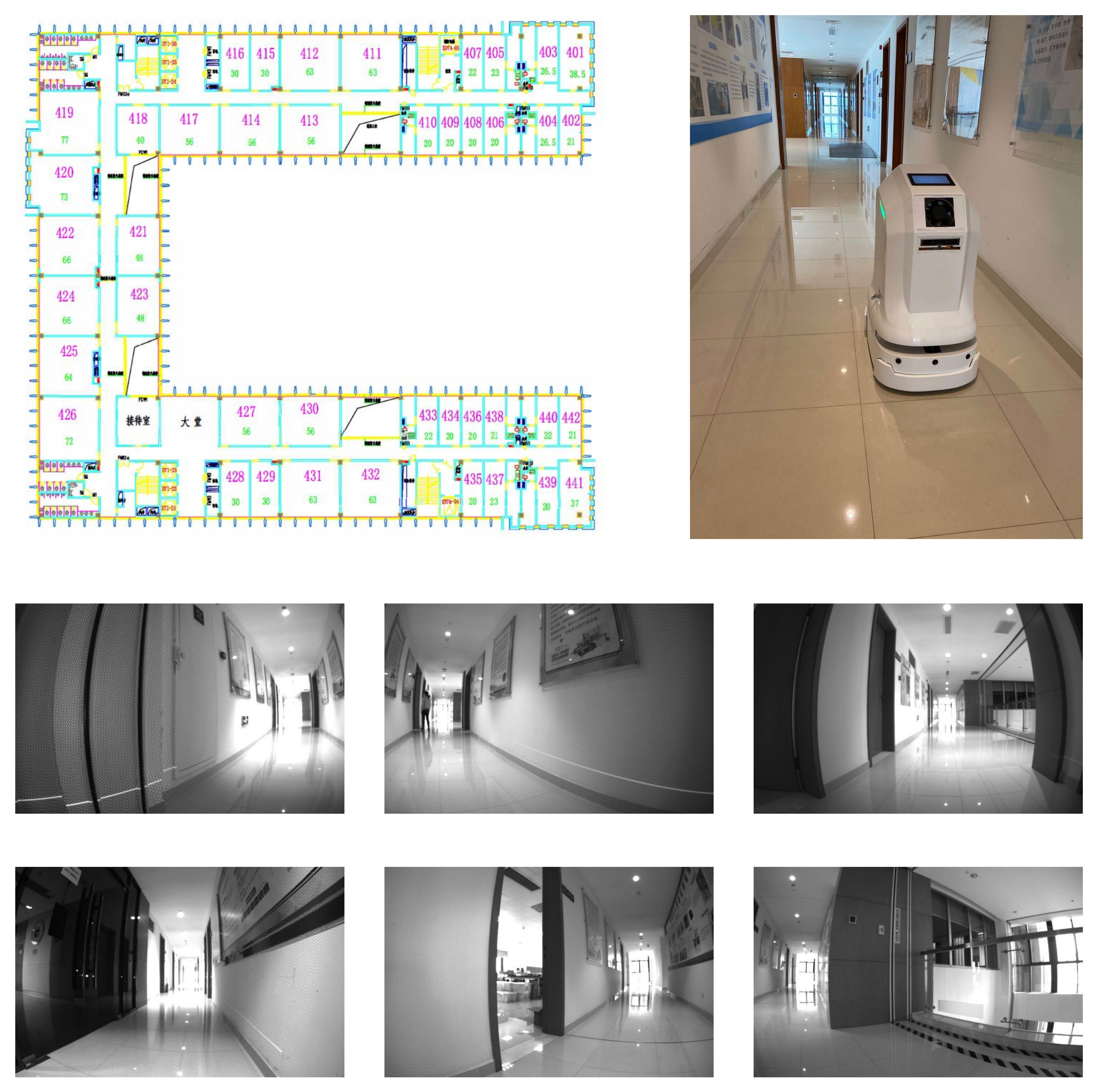}
\caption{The layout of the experimental corridor environment, the mobile robot constructing $\mathcal{DB}_{W}$ and  $\mathcal{DB}_{I}$ automatically and several image samples from $\mathcal{DB}_{I}$.}
\label{fig:environment}
\end{figure}

In this section, we conduct several numerical experiments to show the effectiveness of our proposed system. To be more specific, we compare the proposed scheme with two baseline systems, e.g., {\em Baseline 1}: SVM based WiFi only localization scheme and {\em Baseline 2}: WiFi and vision integrated based scheme in \cite{hu2017wi}. The proposed localization scheme is verified in the corridor environment of an office building with 4000 square meters, where the layout is shown in Fig.~\ref{fig:environment}.

In order to obtain the ground truth positions and establish the databases $\mathcal{DB}_{W}$ and $\mathcal{DB}_{I}$, we have done the following implementation work. Firstly, according to the mobile robot's installation location, the relative position coordinate of camera and IMU module is obtained. Secondly, the real-time coordinates of the mobile robot on the floor are calculated by constructing a two-dimensional floor map through robot SLAM. Thirdly, multiple cruise points are set on map of SLAM, enabling the robot to complete the task while moving autonomously on the floor corridor. Fourthly, the mobile robot take multiple s-shaped trajectory routes to collect data and several slopes are placed in the corridor to increase the position changes on Z-axis. In order to ensure the user localization accuracy, the mobile robot will collect the wireless signal and image data periodically to keep $\mathcal{DB}_{W}$ and $\mathcal{DB}_{I}$ updated. The data training process is conducted on a localization server with NVIDIA Titan X GPU with Pytorch platform and other detailed parameter values are listed in Table~\ref{tab:parameter}. Kindly note that when the offline training process is completed, the online localization service will be provided by the indoor localization server.

\begin{table}
\caption{The detail parameter values of the experiments} \label{tab:parameter}
\Large
\centering
\footnotesize
\begin{tabular}{c c c c}
\toprule
\textbf{Parameter} & \textbf{Value} & \textbf{Parameter} & \textbf{Value} \\
\midrule
$N_s$ & 50 & $N_{AP}$ & 5  \\
\midrule
$N_p$ & 104 & $N_w$  & 752  \\
\midrule
$N_l$ & 780 & $N_{RGB}$ & 1 \\
\midrule
$N_{RP}$ & 24 & $N_{\mathcal{A}}$ & 15 \\
\midrule
$N_{W}$ & 2 & $N_{I}$ & 2 \\
\midrule
$J^{\star}$ & 4 & \\
\bottomrule
\end{tabular}
\end{table}

\subsection{Localization Accuracy}
\begin{figure}
\centering
\includegraphics[width = 3.4 in]{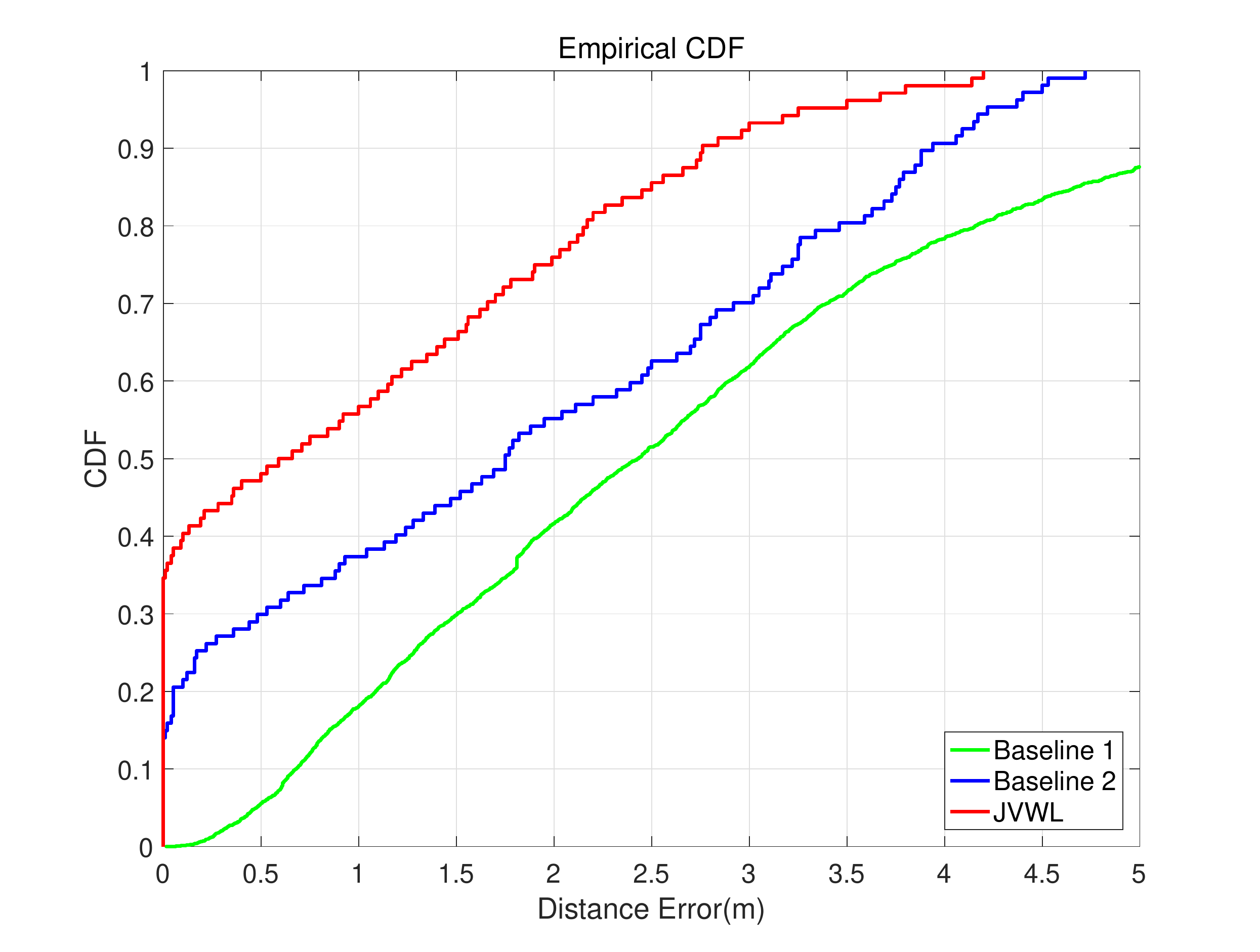}
\caption{CDF of localization distance errors for different algorithms in experimental corridor environment. The proposed method based wireless signal with CNN architectures is compared with two baselines to test the algorithm effectiveness.}
\label{fig:distance error}
\end{figure}

In this experiment, we investigate the localization accuracy in terms of the cumulative distribution function (CDF) of distance errors, by comparing the proposed JVWL scheme with the above two baselines. From Fig.~\ref{fig:distance error}, we can observe that the proposed method (red solid curves) achieves more accurate and reliable localization results than {\em Baseline 1} scheme (green solid curves) as well as {\em Baseline 2} scheme (blue solid curves), considering no extra image sign for feature extraction. Specifically, the proposed JVWL scheme achieves the median localization error of 0.62 m, which is superior to 2.43 m in {\em Baseline 1} scheme or 1.75 m in {\em Baseline 2} scheme. In addition, 57\% cases achieves the localization errors within 1 m for the proposed scheme, while other baseline schemes only achieves less than 40\% cases.

\subsection{Grid Size Effect}
In this experiment, we investigate the effects of grid sizes to find a practical tradeoff between deployment cost and localization accuracy. Based on this tradeoff relation, we can figure out the most efficient deployment strategy for the proposed localization system. To this end, distances between adjacent RPs are set to be 1 m, 1.5 m and 2 m, respectively.

\begin{figure}
\centering
\includegraphics[width = 3 in]{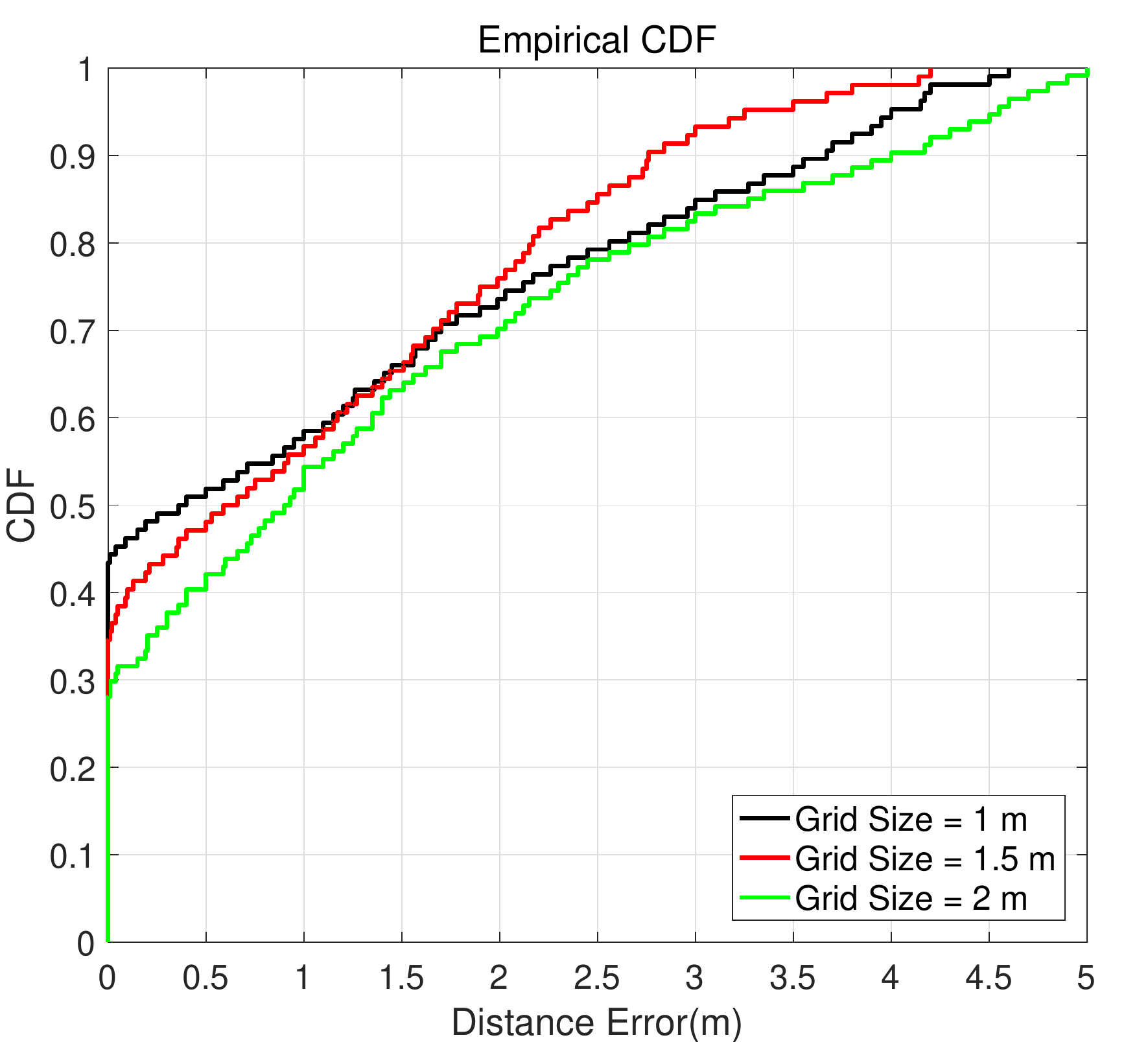}
\caption{CDF of localization errors for different training cell size. Three training datasets with different cell sizes are tested respectively to explore the most efficient deployment method.}
\label{fig:size}
\end{figure}

Localization errors under different grid sizes are illustrated in Fig.~\ref{fig:size}, where the corresponding average localization errors are 0.83 m (black solid curves), 0.62 m (red solid curves) and 1.19 m (green solid curves), respectively. It is worth noting that when the size is changed from 1 m to 1.5 m, the localization accuracy improve 25.3\% at the cost of approximately half of the data collection and labelling. Based on the above results, we believe 1.5 m is a reasonable grid size in the database construction stage.

\subsection{Computational Complexity}
Although in terms of localization errors, the proposed scheme provides a satisfactory performance, the implementation complexity is still open. In this experiment, we show the effectiveness of proposed schemes by comparing the total computational time cost with the above baseline schemes. As shown in Table~\ref{tab:results}, {\em Baseline 1} scheme cost the least time regardless of the sample number, considering only simple matching and classification processes are conducted in the online phase. The average time consumption by the proposed JVWL method is only 0.18s, which is much less than {\em Baseline 2}. This is due to the fact that SITF algorithm in {\em Baseline 2} requires to store feature vectors for each training frame and perform a linear search to find the nearest neighbour for each given test frame, which will cost too much time. The low complexity of the proposed JVWL algorithm make the practical system achieve real-time localization.

\begin{table}[!htbp]
\caption{Total Running Time Comparison for Different Methods.} \label{tab:results}
\centering
\footnotesize
\begin{tabular}{c c c c}
\toprule
\textbf{Test image} & \textbf{Baseline 1} & \textbf{Baseline 2}  & \textbf{JVWL} \\
\midrule
1 & 0.01 s & 1.2 s & 0.18 s  \\
\midrule
10 & 0.02 s & 14.0 s  & 1.8 s  \\
\midrule
100 & 0.1 s & 125.0 s & 19.2 s \\
\bottomrule
\end{tabular}
\end{table}

\section{Conclusion} \label{sect:conc}
In this paper, we propose a joint visual and wireless signal feature based approach for high-precision indoor localization. By applying a coarse and fine localization framework, we formulate the joint localization problem accordingly. Based on theoretical analysis, a deep learning neural network is applied to estimate the fine localization results. Compared with the conventional single technology based scheme, our localization system achieves a high accuracy of about 0.62 m with low computational complexity in the standard corridor environment through numerical experiments.

\section*{Acknowledgement}
This work was supported in part by the National Natural Science Foundation of China (NSFC) Grants under No. 61701293 and No. 61871262, the National Science and Technology Major Project (Grant No. 2018ZX03001009), the National Key Research and Development Program of China (Grant No. 2017YFE0121400), the Huawei Innovation Research Program (HIRP), and research funds from Shanghai Institute for Advanced Communication and Data Science (SICS).

\bibliographystyle{IEEEtran}
\bibliography{IEEEabrv,wifi}
\end{document}